# Anisotropic Scattering and Anomalous Transport in a High Temperature Superconductor

M. Abdel-Jawad[*], M. P. Kennett[+], L. Balicas[$], A. Carrington[*], A. P. Mackenzie[#], R. H. McKenzie[@] & N. E. Hussey[*]

[*]H.H. Wills Physics Laboratory, University of Bristol, Tyndall Avenue, Bristol, BS8 1TL, UK

[+]Physics Department, Simon Fraser University, 8888 University Drive, Burnaby, British Columbia, V5A 1S6, Canada

[$]National High Magnetic Field Laboratory, Florida State University, Tallahassee, Florida, 32306, USA

[#]School of Physics and Astronomy, University of St. Andrews, St. Andrews, Fife, KY16 9SS, UK

[@]Physics Department, University of Queensland, Brisbane 4072, Australia

**The metallic state of high temperature cuprate superconductors is markedly different from that of textbook metals. The origin of this unconventional state, characterized by unusual and distinct temperature dependences in the transport properties[1-4], remains unresolved despite intense theoretical efforts.[5-11] Our understanding is impaired by our inability to determine experimentally the temperature and momentum dependence of the transport scattering rate. Here we use a novel magnetotransport probe to show that the unusual temperature dependences of the resistivity and the Hall coefficient in highly doped $Tl_2Ba_2CuO_{6+\delta}$ originate from two distinct inelastic scattering channels. One channel is due to conventional electron-electron scattering whilst the other is highly anisotropic, has the same symmetry as the superconducting gap and a magnitude that grows approximately linearly with temperature. The observed form and anisotropy put tight constraints on theories of the metallic state. Moreover, in heavily doped *non-***

**superconducting cuprates, this anisotropic scattering term appears to be absent[12], suggesting an intimate connection between the origin of this scattering and superconductivity itself.**

The in-plane properties of layered metals can sometimes be obtained from measurements of out-of-plane quantities. For example, angular magnetoresistance oscillations (AMRO), angular variations in the interlayer resistivity $\rho_\perp$ induced by rotating a magnetic field $H$ in a polar plane relative to the conducting layers, have provided detailed information on the shape of in-plane Fermi surface (FS) in a variety of layered metals, including the overdoped cuprate $Tl_2Ba_2(Ca_0)Cu_1O_{6+\delta}$ (Tl2201).[13] Here we resolve for the first time the momentum ($k$) and energy ($\omega$ or $T$) dependence of the transport lifetime $\tau$ in cuprates through advances, both experimental and theoretical, in the AMRO technique. Experimentally, we extend the temperature range of measurements on overdoped Tl2201 (with a superconducting transition temperature $T_c$ = 15K) by more than one order of magnitude. Theoretically, we derive a new general analytical expression for the interlayer conductivity $\sigma_\perp$ in a tilted $H$ which incorporates basal-plane anisotropy. For $T > 4$K, the AMRO can only be explained by an anisotropic scattering rate $1/\tau$ whose anisotropy grows with $T$. Significantly, the anisotropy in $1/\tau$ and its $T$-dependence up to 55K can quantitatively account for both the robust linear-in-$T$ component to the in-plane resistivity $\rho_{ab}$ and the $T$-dependent Hall coefficient $R_H$ over the same temperature range.[14,15] These anomalous behaviours are not characteristics of a simple Fermi liquid, often the starting point for modelling overdoped cuprates. We discuss the consequences of these findings for our understanding of the normal state transport and the pairing interaction.

As described in the Supplementary section, detailed azimuthal and polar angle dependent AMRO data were taken at 4.2K and 45 Tesla and fitted to the Shockley-Chambers tube integral form of the Boltzmann transport equation, modified for a quasi-2D metal[16] (and assuming an isotropic mean-free-path $\ell$), to generate a full three-dimensional parameterization of the FS, $k_F(\varphi,\theta)$, consistent with previous measurements[13]. Before studying the *T*-dependence of the scattering rate, a self-consistency check was carried out on the fitting procedure by varying **H** at a fixed temperature. The solid lines in Fig. 1a) represent polar angle dependent $\Delta\rho_\perp(\theta)/\rho_\perp(H=0)$ data at 4.2K (normalized to the zero-field resistivity) for various fields $20T \leq \mu_0H \leq 45T$ at a fixed azimuthal orientation of the inclined sample $\phi = 29°$ (relative to the Cu-O-Cu bond direction) where all AMRO features are visible. The magnetoresistance is determined by the magnitude of $\omega_c\tau$ where $\omega_c$ is the cyclotron frequency. The dashed lines are simulated $\Delta\rho_\perp(\theta,\phi=29°)/\rho_\perp$ curves produced simply by scaling $\omega_c\tau$ (= $0.41(\mu_0H/45)$) *whilst keeping all other parameters fixed at their 45T values*. The data scale very well, implying that the isotropic formalism[16] remains valid with decreasing *H* and that no additional angular dependence appears due to the presence of inhomogeneous superconducting regions (with different $T_c$ values) or anomalous vortex liquid phases.[17]

Fig. 1b) shows the temperature dependence of $\Delta\rho_\perp(\theta,\phi=29°)/\rho_\perp$ up to 55K ($\mu_0H$ = 45T). Remarkably, AMRO features remain discernible at all temperatures, in particular the kink around $\theta = 45°$. However, comparison of the data in Fig. 1a) and 1b) reveals that the AMRO evolve differently depending upon whether $\omega_c\tau$ is reduced by decreasing *H* or by increasing *T*. In the former case, both the peak at **H**//**c** and the peak at intermediate angles diminish at approximately the same rate, whilst in the latter, the intermediate peak is found to survive up to much higher temperatures. The dashed lines in Fig. 1b) show the best least-squares fits to the data *assuming all parameters except the product $\omega_c\tau$ remain constant up to 55K*. These fits are clearly inferior to those in Fig. 1a).

To proceed, we relax the constraint that $\omega_c\tau$ remains isotropic at all temperatures and generalize the expression for the interlayer conductivity $\sigma_\perp$[16] to incorporate basal plane anisotropy in the relevant parameters. We first define the Fermi velocity as $\mathbf{v}_F(\varphi) = v_F^0(1+\beta\cos4\varphi)$ and the variation of $\omega_c$ around the FS as

$$\omega_c(\varphi,\theta) = e\mu_0 H \cos\theta \frac{\mathbf{k}_F(\varphi)\cdot\mathbf{v}_F(\varphi)}{\hbar k_F(\varphi)^2} \quad (1)$$

The generalized expression for $\sigma_\perp$ then becomes

$$\sigma_\perp = \frac{e^2}{4\pi^3\hbar^2}\frac{1}{1-P}\int_{-\pi/d}^{\pi/d}dk_\perp \int_0^{2\pi}d\varphi_2 \frac{e\mu_0 H\cos\theta}{\omega_c(\varphi_2)}\int_{\varphi_2-2\pi}^{\varphi_2}d\varphi_1\, v_\perp(\varphi_2,k_\perp)v_\perp(\varphi_1,k_\perp)\frac{G(\varphi_2,\varphi_1)}{\omega_c(\varphi_1)} \quad (2)$$

where $k_\perp$ is the *c*-axis reciprocal lattice vector, $v_\perp$ the interlayer velocity, $d$ the interlayer spacing (= 1.16 nm for Tl2201), $P = G(2\pi, 0)$ is the probability that an electron makes a complete orbit of the FS without being scattered and

$$G(\varphi_2,\varphi_1) = \exp\left(-\int_{\varphi_1}^{\varphi_2}\frac{d\varphi}{\omega_c(\varphi)\tau(\varphi)}\right) \quad (3)$$

This formalism holds irrespective of whether hopping is coherent or weakly incoherent (i.e. when $\hbar/\tau > 2t_\perp$ and $v_\perp$ is ill-defined).[18] In the latter case, AMRO arise from differences in Aharonov-Bohm phases acquired in hopping between layers for positions $\varphi_1$ and $\varphi_2$ on the FS.[19]

Consistent with the tetragonal symmetry of Tl2201, we define $1/\tau(\varphi) = (1 + \alpha \cos 4\varphi)/\tau_0$. Whilst no unique and independent determination of the various anisotropy parameters can be made from fits of theoretical curves to AMRO data alone, certain features of the data tightly constrain the parametrization, in particular the FS parameters defining $k_F(\varphi,\theta)$. Furthermore, as there is no experimental evidence to suggest changes in the FS topography with temperature, we fix these parameters to their values at 4.2K. Similarly, $\beta$, the anisotropy in $v_F$, is assumed to be constant. Finally, in order to minimise the number of fitting parameters, we assume that $\omega_c$ is isotropic ($= \omega_0$) within the basal plane. Thus we can provisionally ascribe the evolution of the AMRO uniquely to changes in $1/\omega_0\tau(\varphi)$ and extract $1/\omega_0\tau_0(T)$ and $\alpha(T)$ from fits to the data at different temperatures. The best least-square fits are shown in Fig. 1c). The quality of the fits at all $T$ is clearly much improved with just the inclusion of $\alpha(T)$. Moreover, the subsequent fitting to the in-plane transport data is sufficiently good (see below) that the introduction of additional parameter(s), e.g. to account for any possible $T$-dependence in $\beta$, appears unnecessary.

The consequences of the above analysis are examined in Fig. 2. To aid our discussion, we show schematically in Fig. 2a) the in-plane geometry of various relevant entities with respect to the 2D projection of the FS of overdoped Tl2201 (red curve in Fig. 2a)). The purple line represents the *d*-wave superconducting gap whilst the blue solid line depicts our deduced geometry of $1/\omega_c\tau(\varphi)$ (as governed by the sign of $\alpha$), its maximum being at $\varphi = 0°$. Note that the scattering anisotropy and the superconducting order parameter have the same symmetry. This is consistent with earlier *azimuthal* AMRO data[20] but contrasts with recent angle-resolved photoemission spectroscopy (ARPES) measurements.[21] We note however that the ARPES-derived scattering rate is one order of magnitude larger, suggesting that the two probes are not measuring the same quantity.

In order to give our anisotropic function for $\omega_c\tau$ more physical meaning, we re-express $(1 + \alpha\cos4\varphi)/\omega_0\tau_0$ as $(1-\alpha)/\omega_0\tau_0 + (2\alpha/\omega_0\tau_0)\cos^2 2\varphi$. The *isotropic* part $(1-\alpha)/\omega_0\tau_0$ (black dashed line in Fig. 2a)) is the sole contribution along the diagonal 'nodal' direction (indicated by the green arrow) where the pairing gap vanishes. The *T*-dependence of $(1-\alpha)/\omega_0\tau_0$ is plotted in Fig. 2b) and as shown by the dashed line, follows a simple quadratic law $(A + BT^2)$. By contrast, the *anisotropic* component $2\alpha/\omega_0\tau_0$, maximal in the direction given by the orange arrow in Fig. 2a) and plotted in Fig. 2c), is seen to grow approximately linearly with temperature, *this linearity extending at least down to 4.2K*.

To our knowledge, this is the first quantitative determination of the momentum and temperature dependence of the in-plane mean-free-path in cuprates. Together with the complete FS topology, this is all we need to calculate various coefficients of the in-plane conductivity tensor. Fig. 2d) shows $\rho_{ab}(T)$ as determined from our analysis, superimposed on published data for overdoped Tl2201 at the same doping level (with the superconductivity suppressed by a large magnetic field).[14] The form of $\rho_{ab}(T)$, in particular the strong *T*-linear component below 10K and the development of supra-linear behaviour above this temperature, is extremely well reproduced by the model. The corresponding $R_H(T)$ is shown in Fig. 2e). Significantly, *the absolute change in anisotropy in* $(\omega_0\tau)^{-1}(\varphi, T)$ can account fully for the rise in $R_H(T)$, at least up to 40K. Above 40K, the simulation has a slightly weaker *T*-dependence, possibly due to the increased disorder in the AMRO sample, known to weaken the overall *T*-dependence of $R_H(T)$ in cuprates,[2] and/or the emergence of vertex corrections that manifest themselves only in the in-plane transport.[22] Overall however, the same parametrization of $1/\omega_0\tau(\varphi,T)$ described in Fig. 2b) and 2c) gives an excellent account, not only of the evolution of the AMRO signal (Fig. 1c)), but also of the 'anomalous' transport behaviour. Given the gradual evolution of the transport properties in Tl2201 with doping,[23] we believe these findings will be relevant to crystals with higher $T_c$ values.

We now turn to discuss the implications of our results for existing theories of transport in high-$T_c$ cuprates. Several contrasting approaches dominate much current thinking; Anderson's resonant-valence-bond picture,[24] marginal Fermi-liquid phenomenology[6] and models based on fermionic quasiparticles that invoke specific (anisotropic) scattering mechanisms within the basal plane due either to anisotropic electron-electron (possibly Umklapp) scattering[7] or coupling to a singular bosonic mode, be that of spin,[8,9] charge[10] or superconducting fluctuations.[11] Our analysis clearly supports the latter models in which anisotropy in the *inelastic* part of $\ell(k)$ is responsible for the anomalous $R_H(T)$. Both $R_H(T)$ and the *T*-linear component of $\rho_{ab}(T)$ are derived from a *T*-linear anisotropic scattering term that is maximal along the Cu-O-Cu bond direction. The magnitude of the anisotropy is large, even at such an elevated doping level. At $T = 55$K, for example, $\ell(\boldsymbol{k})$ varies by a factor of two around the in-plane FS. Significantly, in non-superconducting cuprates, $\rho_{ab}(T) \propto T^2$ at low temperatures with no evidence of a $T$-linear term.[12,23] This implies that the development of superconductivity (from the overdoped side) is closely correlated with the appearance of the $T$-linear resistivity and anisotropic inelastic scattering. (Recall that $1/\tau(\varphi)$ also has the same angular dependence as the superconducting gap.)

Our analysis implies the presence of (at least) two inelastic scattering channels in the current response of superconducting cuprates. Recent ARPES measurements[25] on $Bi_2Sr_2CaCu_2O_{8-\delta}$ also found evidence for two contributions to the *quasiparticle* scattering rate; one quadratic and one linear in $\omega$ that develops a kink below $T_c$. A scattering process that is quadratic in both temperature *and* frequency is characteristic of electron-electron scattering. Given that Hall conductivity is dominated by those regions (in this case, the nodal regions) where scattering is weakest, we tentatively ascribe the $T^2$ dependence of the Hall angle $\cot\theta_H$ in cuprates to such scattering.

The second term (seen by ARPES) has been attributed to scattering off a bosonic mode, though its origin and its relevance to high-$T_c$ superconductivity remain subjects of intense debate.[26] Possible candidates include phonons, $d$-wave pairing fluctuations, spin (large-**q**) fluctuations and charge (small-**q**) fluctuations but since all, bar phonons, appear to vanish in heavily overdoped non-superconducting cuprates,[11,27,28] it is difficult to single one out at this stage. Nevertheless, if this bosonic mode is the source of the anisotropic scattering revealed by AMRO, the continuation of its linear $T$-dependence to very low temperatures implies the presence of a surprisingly low energy scale. Whatever its origin, the apparent correlation between superconductivity and the anomalous scattering makes its resolution a prime route to identify the pairing mechanism for high-$T_c$ superconductivity and the form of this anomalous scattering, at least in the normal state, has now been identified. It is also worth considering whether the anisotropic scattering reported here is a remnant of the more intense scattering found in the underdoped regime where checkerboard charge order develops.[29] Recall that $\cot\theta_H$ does not vary markedly across the cuprate phase diagram and so the strength of electron-electron scattering appears largely doping independent. By contrast, the stronger $T$-linear behaviour seen in $\rho_{ab}$ as one approaches maximum $T_c$ points to an increase in the anomalous term with lower doping. Below optimal doping, the strength of anisotropic scattering will continue to grow as one approaches the Mott insulator, driving the simple anisotropic metal into a more exotic 'nodal' metallic state in which the FS is reduced to a series of Fermi arcs in those (nodal) regions where scattering is weakest.[30] Clearly, the connection between the anisotropy in the under- and overdoped regimes is an important avenue for future research.

Finally, this work demonstrates that AMRO can be an extremely powerful probe of intra-layer anisotropies in layered metals, beyond mere determination of the FS. The formalism and procedure we have employed here could be applied to a host of other layered correlated metals, e.g. molecular superconductors and ruthenates, to establish whether anisotropic scattering also plays an important role in the unconventional behaviour in these systems.

**Acknowledgements** We thank J. C. Davis, L. P. Gor'kov, B. L. Gyorrfy, P. B. Littlewood, A. J. Schofield, N. Shannon and J. A. Wilson for helpful discussions. We also acknowledge assistance from V. Williams in construction of the two-axis rotator, designed and conceived by LB. This work was supported by EPSRC and a co-operative agreement between the State of Florida and NSF. The crystals were supplied by AC and APM, experiments carried out by MA-J, LB and NEH and the analysis performed by MA-J, MPK, RHM and NEH.

**Competing interests statement** The authors declare that they have no competing financial interests.

**Correspondence** and requests for materials should be addressed to N.E.H. (n.e.hussey@bristol.ac.uk)

**FIGURE LEGENDS**

**Figure 1**. Field and temperature dependencies of the polar AMRO in overdoped Tl2201 at a fixed azimuthal direction $\phi = 29^0$ (the black curve in Fig. 1b)). (a) Solid lines: normalized $\Delta\rho_\perp(\theta,\phi=29^0)$ data for different field strengths as indicated. Dashed lines: simulated AMRO fits using the same $k_{mn}$ coefficients as given in Fig. S1 of the Supplementary section and $\omega_c\tau$ values scaled simply by the field scale (i.e. $\omega_c\tau = 0.41(\mu_0H/45)$). (b) Solid lines: normalized $\Delta\rho_\perp(\theta,\phi=29^0)$ data at 45 Tesla for different temperatures between 4.2K and 55K. Dashed lines: best least-squares fits using the same $k_{mn}$ coefficients as given in Fig. S1 of the Supplementary section and assuming an isotropic $\omega_c\tau$. (c) As (b) but now with an *anisotropic* $\omega_c\tau = \omega_0\tau_0/[1 + \alpha\cos4\varphi]$.

**Figure 2**. Determination of the in-plane transport coefficients from 45 Tesla polar AMRO. (a) Red curve: schematic 2D projection of the FS of overdoped Tl2201. Purple curve: schematic representation of the *d*-wave superconducting gap. Blue curve: geometry of $(\omega_c\tau)^{-1}(\varphi)$. Black dashed line: isotropic part of $(\omega_c\tau)^{-1}(\varphi)$. (b) *T*-dependence of $(1-\alpha)/\omega_0\tau_0$, i.e. the isotropic component of $(\omega_c\tau)^{-1}(\varphi)$ and sole contribution along the 'nodal' region indicated by the green arrow in 3(a). The green curve is a fit to $A + BT^2$. (c) *T*-dependence of $2\alpha/\omega_0\tau_0$, i.e. the anisotropic component of $\omega_c\tau^{-1}(\varphi)$ and the additional contribution that is maximal along the 'anti-nodal' direction indicated by the orange arrow in 3(a). The orange curve is a fit to $C + DT$. (d) Black circles: $\rho_{ab}(T)$ data for overdoped Tl2201 ($T_c$ = 15K)

extracted from Ref. [14]. Purple curve: simulation of $\rho_{ab}(T)$ from parameters extracted from our AMRO analysis. To aid comparison, 1.9$\mu\Omega$cm have been subtracted from the simulated data. (It is not unreasonable to expect different crystals to have different residual resistivities.) (e) Black circles: $R_H(T)$ data for the same crystal [14]. Purple curve: simulation of $R_H(T)$ from parameters extracted from our AMRO analysis. In this case, no adjustments have been made. $\rho_{ab}(T)$ and $R_H(T)$ were calculated using the Jones-Zener expansion of the linearized Boltzmann transport equation for a quasi-two-dimensional Fermi surface.[7] Note that using (1), we can re-express the expressions in Ref. [7] solely in terms of parameters extracted from our analysis.

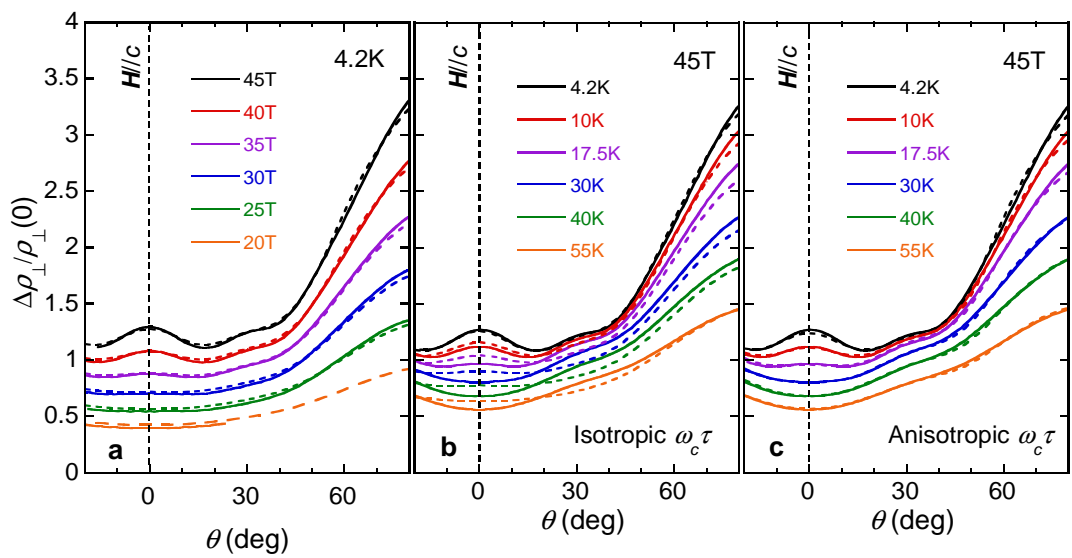

Figure 1
Abdel-Jawad *et al*.

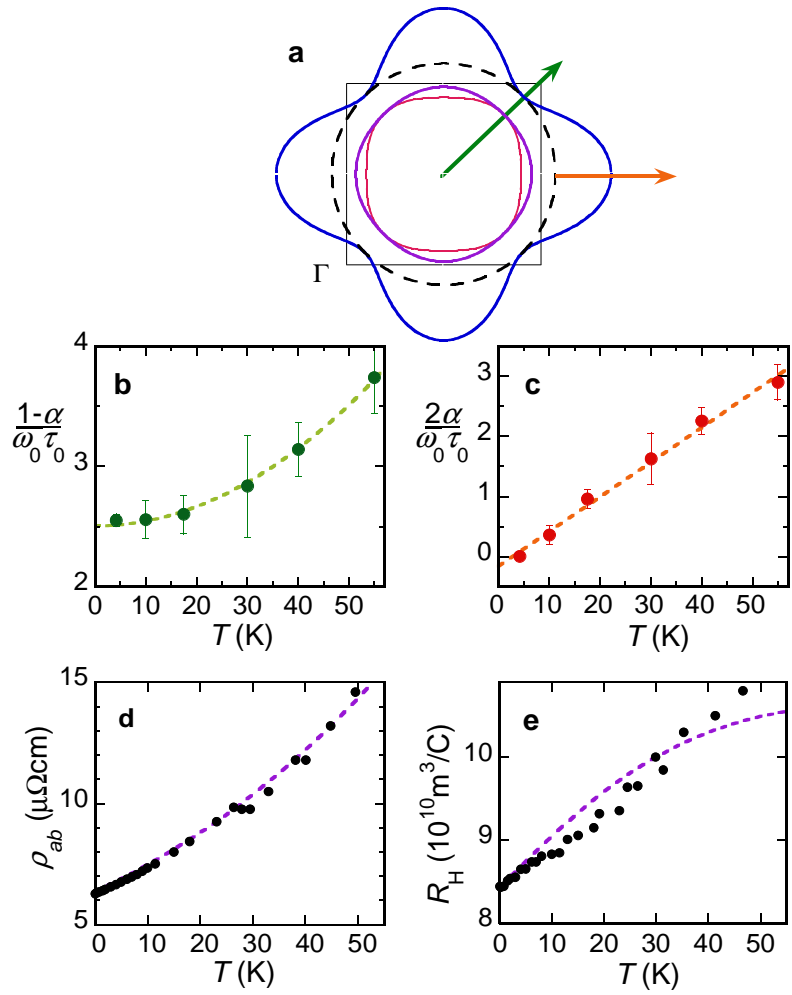

Figure 2
Abdel-Jawad *et al.*

**Supplementary Materials**

**Determination of the Fermi surface parameters from low-temperature AMRO**

Supplementary Fig. S1a) shows normalized $\Delta\rho_\perp(\phi, \theta)/\rho_\perp(H=0)$ data for overdoped Tl2201 ($T_c$ = 15K) taken at 4.2K and 45T. (For experimental details see Ref. [13].) The different coloured traces depict polar AMRO sweeps (i.e. as a function of $\theta$) taken at given azimuthal angles $\phi$ relative to the basal Cu-O-Cu bond direction ([100] or $\{k_x, k_y\}$ = ($\pi/a$, 0)). The key features of the data are very similar to those reported earlier on a crystal of similar $T_c$ over a larger $\phi$ range.[13] In order to fit the AMRO data, we apply the Shockley-Chambers tube integral form of the Boltzmann transport equation, modified for a quasi-2D metal;[16]

$$\sigma_\perp = \frac{e^2}{4\pi^3\hbar^2}\frac{1}{1-P}\int_{-\pi/d}^{\pi/d}\cos\theta\, dk_\perp \int_0^{2\pi}d\varphi_2 \int_{\varphi_2-2\pi}^{\varphi_2}d\varphi_1\left\{v_\perp(\varphi_2,k_\perp)v_\perp(\varphi_1,k_\perp)\frac{m_c^*}{\omega_c}G(\varphi_2-\varphi_1)\right\} \quad (1)$$

Here $G(\varphi_2, \varphi_1) = e^{-(\varphi_2-\varphi_1)/\omega_c\tau}$. The other parameters are defined in the main text. Note that the formula involves correlations in $v_\perp$ at different times (or equivalently different $\varphi$). For a simple warped cylindrical FS, the energy dispersion is $E(\mathbf{k}) = (\hbar^2/2m^*)k_F^2 - 2t_\perp\cos(k_\perp d)$ where $2t_\perp$ ($= \hbar v_\perp/d$) is the interlayer hopping energy, $k_\perp = k_{\perp 0} - k_F\cos\phi\tan\theta$, and $k_{\perp 0}$ serves as an index of the FS perpendicular to the field direction. If all other parameters in (1) are isotropic within the basal plane then clearly $\sigma_\perp$ has no azimuthal dependence. In order to account for such strong azimuthal dependence, one typically includes just anisotropy of the FS geometry in (1). To a first approximation, Tl2201 has a body-centred tetragonal structure and the simplest *anisotropic* dispersion that respects the appropriate symmetry constraints and fits the data is $E = (\hbar^2/2m^*)k_F^2(\varphi) - 2t_\perp \cos(k_\perp d)(\sin2\varphi + k_{61}\sin6\varphi + k_{101}\sin10\varphi)$ where $k_F(\varphi) = k_{00}(1 + k_{40}\cos4\varphi)$ and $k_{40}$, $k_{61}$ and $k_{101}$ are anisotropy parameters. Note that all three components in the *c*-axis warping are required to fit the data.[13]

Supplementary Fig. S1b) shows the best resultant fits to (1) with the parameters given in the captions. All aspects of the data, including the crossing point at $\phi = 55^\circ$, are reproduced. The good quality of fit implies that other anisotropies not included in the fitting, e.g. in the lifetime or velocity, are relatively small at 4.2K and therefore the quasi-particle mean-free-path $\ell$ is approximately isotropic in the 'impurity-dominated' regime at low *T*. Whilst we have no *a priori* reason to expect the isotropic-$\ell$ approximation to be applicable, this observation explains the good consistency between the measured value of $R_H(T=0)$[14] and the size of the in-plane FS determined independently by AMRO[13] and ARPES.[19] Supplementary Fig. S1c) shows a three-dimensional representation of the FS derived from the same parameters. These parameters are used for all subsequent analysis.

**Procedure to fit AMRO with anisotropic $\omega_c$**

The procedure used to fit AMRO in Figs S1b, 1a-c, and to determine Figs 2b) and 2c) contains the assumption that there is anisotropy in $k_F(\varphi)$, but none in $\omega_c$. This assumption implies that the anisotropies in $v_F(\varphi)$ and $k_F(\varphi)$ are equivalent and small, i.e. $\beta = k_{40} << 1$. Generically, fourfold anisotropy in $k_F(\varphi)$ will be accompanied by fourfold anisotropy in $\omega_c(\varphi)$ via (2), and hence the simplest self-consistent approach to fitting AMRO is to include anisotropy in $\omega_c(\varphi)$. We do this by writing $1/\omega_c(\varphi) = 1/\omega_0 (1 + u\cos4\varphi)$ (and thereby adding one extra parameter to the fitting routine). The tight binding dispersion determined by ARPES for a Tl2201 sample with a similar $T_c$ to the sample studied here would suggest that $u > 3k_{40}$.[21] Fits to the AMRO here are consistent with this approximate relation, with $u \sim -0.14$.

Examination of equations (2) and (3) in the main text shows that the product $\omega_c(\varphi)\tau(\varphi)$ enters the expression for $\sigma_\perp$ as the argument of an exponential and hence is quite sensitive to changes in the anisotropy of either quantity. In particular, we observe that $\alpha(T)$ as shown in Fig 2c) is actually shifted from the $\alpha(T)$ determined when $u$ is non-zero by the approximate relation $\alpha_{u=0}(T) \sim \alpha(T) + u$. As $\alpha(T)$ is no larger than about 0.3, this implies that whilst $\alpha_{u=0}(T)$ has a qualitatively correct temperature dependence, quantitative determination of $\alpha(T)$ will in general require fits with $u$ not equal to 0. This is illustrated in Supplementary Fig. S2. It should be emphasised that although the neglect of anisotropy in $\omega_c$ leads to shifted determinations of $\alpha(T)$, the observation of anisotropy in the scattering and its growth with temperature is robust, and assuming anisotropic $\omega_c(\varphi)$ leads to no quantitative improvement to the fits in Fig. 1c).

Finally, the ARPES tight-binding parametrisation of the dispersion[21] actually also suggests that $\cos8\varphi$ terms should be present in $k_F(\varphi)$ and $1/\omega_c(\varphi)$. Fits involving these extra parameters (and even a $\cos8\varphi$ term in $\tau(\varphi)$) leave the coefficients of the $\cos4\varphi$ terms relatively unchanged, confirming the robustness of the fitting procedure to determine anisotropy in the scattering rate.

**Supplementary Figure S1. Parameterization of the Fermi surface from AMRO data.**

**a,** AMRO sweeps in polar angle $\theta$ at various azimuthal angles $\phi$ for an overdoped Tl2201 single crystal ($T_c$ = 15K). The data were taken at $T$ = 4.2K and $H$ = 45T and are normalized to the zero-field value of $\rho_\perp$. The different azimuthal orientations ($\pm$ 1°) of each polar sweep are stated relative to the basal Cu-O-Cu bond direction. **b,** Best least-squares fit to the AMRO data with $k_F(\phi, k_\perp) = k_{00}(1 + k_{40}\cos4\varphi) + 2t_\perp \cos(k_\perp d)(\sin2\varphi + k_{61}\sin6\varphi + k_{101}\sin10\varphi)$. Here $\omega_c\tau$ = 0.41(0.03), $k_{00}$ = 7.33(0.1)nm$^{-1}$, $k_{40}$ = - 0.047(0.004), $k_{61}$ = 0.68(0.06) and $k_{101}$ = -0.2(0.05). These values are in good agreement with those obtained previously on a second crystal [13]. Note that AMRO themselves are not dependent on the absolute value of $2t_\perp$. **c,** Reconstruction of the FS in Tl2201 from the polar AMRO data. The magnitude of the c-axis warping has been increased for emphasis.

**Supplementary Figure S2, Comparison of anisotropy parameters extracted from isotropic- and anisotropic-$\omega_c$ approximations.**

**a,** $T$-dependence of $(1-\alpha)/\omega_0\tau_0$, i.e. the isotropic component of $\omega_c\tau^{-1}(\varphi)$ for different parametrizations. The green closed circles are the same parameters shown in Fig. 2b) assuming $\omega_c$ is isotropic. The green squares are the corresponding parameters assuming an anisotropic $\omega_c(\varphi) = \omega_0/(1 + u\cos4\varphi)$ with $u$ = -0.14 determined from a fit to the data at $T$ = 4.2K. This value agrees favourably with that expected from the ARPES results of Plate *et al.* [21]. The dashed curves are fits to $A + BT^2$. **b,** *T*-dependence of $2a/\omega_0\tau_0$, i.e. the anisotropic component of $\omega_c\tau^{-1}(\varphi)$. The red closed circles are the same parameters shown in Fig. 2c). The red squares are the corresponding parameters assuming an anisotropic $\omega_c(\varphi) = \omega_0/(1 + u\cos4\varphi)$ as in **a,**. The dashed curves are fits to $C + DT$. It should be noted that $D$ is comparable in both fits.

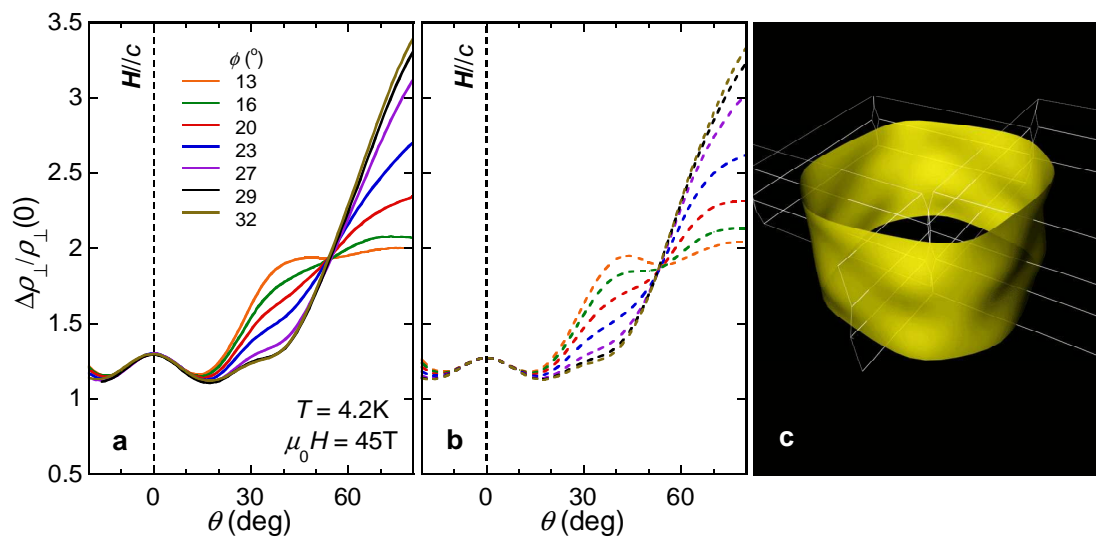

Supplementary Figure 1
M. Abdel-Jawad *et al*.

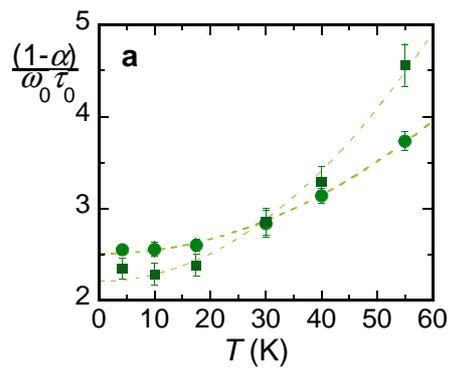
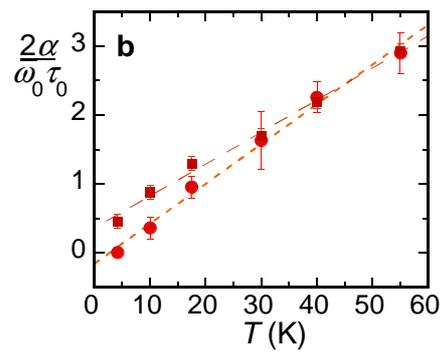

Supplementary Figure 2
M. Abdel-Jawad *et al*.